 \documentclass[12pt]{article}
\usepackage{siunitx}
\usepackage{times} 
\usepackage{amsmath}
\usepackage[T1]{fontenc}
\usepackage[thinc]{esdiff}
\usepackage{array,booktabs}
\usepackage{lipsum} 
\usepackage{amsfonts, amsmath, amsthm, amssymb} 
\usepackage{graphicx} 
\usepackage{booktabs} 
\usepackage{wrapfig} 
\usepackage[top=1in,bottom=1in,left=1in,right=1in]{geometry} 
\usepackage{mathtools}
\usepackage{amsmath}

\usepackage{placeins}
\usepackage{float}
\usepackage{authblk}

\usepackage{graphicx} 
\usepackage[]{caption} 
\usepackage [labelfont=bf]{subcaption}  
\usepackage[raggedleft]{sidecap}   
\DeclareCaptionStyle{sidecap}{position=top}%
\usepackage[center]{titlesec}  
\usepackage[utf8]{inputenc}
\usepackage{xcolor} 

\usepackage{fancyhdr} 
\fancyheadoffset{0cm}

\fancypagestyle{plain}{%
	\fancyhf{}%
	\fancyhead[R]{\thepage}%
}
\providecommand{\keywords}[1]{\textbf{\textit{Keywords: }} #1} 
\begin{document}
	
	\author[1,2]{\fontsize{11}{13}\selectfont Watson Levens}
	\author[3]{Alex Szorkovszky}
	\author[1]{David J. T. Sumpter}
	\affil[1]{ \textit { Department of Information Technology, Uppsala University, Sweden}}
	\affil[2]{ \textit {Department of Mathematics, University of Dar es Salaam, Tanzania} }
	\affil[3]{ \textit {Department of Informatics and RITMO, University of Oslo, Norway} }
		\title{Friend of a friend models of network growth}
		
	\date{\vspace{-6.8ex}} 
	\maketitle
	\thispagestyle{empty} 
	\begin{abstract}	
 One of the best-known models in network science is preferential attachment. In this model, the probability of attaching to a node depends on the degree of {\it all} nodes in the population, and thus depends on global information. In many biological, physical, and social systems, however, interactions between individuals depend only on local information. Here, we investigate a truly local model of network formation – based on the idea of a friend of a friend – with the following rule: individuals choose one node at random and link to it with probability $p$, then they choose a neighbour of that node and link with probability $q$. Our model produces power laws with empirical exponents ranging from $1.5$ upwards and clustering co-efficients ranging from 0 up to $0.5$ (consistent with many real networks). For small $p$ and $q=1$, the model produces super-hub networks, and we prove that for $p=0$ and $q=1$, the proportion of non-hubs tends to $1$ as the network grows. We show that power-law degree distributions, small world clustering and super-hub networks are all outcomes of this, more general, yet conceptually simple model.

\end{abstract}
	
\keywords{Networks, power-laws, degree distributions, clustering coefficients}

\section{Introduction}

One of the important breakthroughs of network science at the start of the millennium was to formulate several models of network growth based on simple assumptions on how individuals connect. For example, Barab\'asi and Albert used the assumption of preferential attachment --- that individuals attach to or follow other individuals with a probability that increases with the number of links (or followers) they already have --- to show that such a behavior results in a power-law distribution of the number of attachments \cite{BA}. This distribution is long-tailed, which means that a few individuals become extremely popular, but not because they are intrinsically more attractive to attach to. Power-law distributions describe substantial inequalities between individuals, as a consequence of the rich getting richer \cite{Adamic, Yakovenko}. There are several empirical examples of networks following power-law distributions  \cite{AlbertB, Clauset, Newman, Barabasi, Pastor, Gabaix, Huberman, Sadri, Gupta, Zang, Nguyen}.

Another well-studied example is the Watts and Strogatz model, which assumes that individuals are connected to a cluster of few local neighbours but then, through a process of rewiring, longer distance connections are made across the network \cite{WattsStrogatz}. Their model reproduces both the clustering seen in real networks --- such that if A knows B and B knows C, then A also knows C --- and the small world effect, whereby any member of the network is only a small number of contacts away from any other person. This average degree of separation was 4.7 for Facebook in 2011 \cite {Ugander}, 4.17 for Twitter in 2012 \cite {Myers}, and 6.6 for the Microsoft instant messaging system in 2006 \cite {Leskovec}.

The Watts and Strogatz model does not give rise to a power-law distribution. So, in order to provide a model that reproduced the power-laws, clustering, and low degree of separation, Holme and Kim \cite{Holme} introduced a further model where the preferential attachment is combined with triad formation, which could produce all three properties. The model proposed by Holme and Kim generates networks with similar properties to the BA model, but can also be tuned to produce clustering similar to that observed in real social networks. Holme and Kim model \cite{Holme} uses triadic closure, previously proposed by Dorogovtsev and his colleagues \cite {Dorogovtsev}, in which a new node randomly selects an edge and then connects with both ends of the edge. This model does not require preferential attachment but does produce a power-law degree distribution, such that the probability that a node has degree $k$ is proportional to $k^{-3}$. A fair amount of research has been done on the mathematical properties of networks based on preferential attachment \cite{BA} and other formation mechanisms \cite{Newman,Carlson,Bianconi, Mitzenmacher} and empirical measurements of real-world networks \cite{ Huberman, Albert, Vazquez}. There also remains some lively debate about whether real-life networks are indeed power-law \cite{Broido, HolmeP,Liu, Voitalov, Serafino}. 

Preferential attachment is an attractive idea since it implies a mechanism whereby individuals follow other people who are more popular than they are. Likewise, Watts and Strogatz rewiring echos how many people's social lives revolve around one group of people who have connections to other groups. In neither case, though, can these models be considered local. The probability of attaching to a node in Barab\'asi and Albert  depends on a proportion, normalized by the sum of the degrees of all members of the population. The same point applied to the Holme and Kim model \cite {Holme}. Likewise, the Watts and Strogatz model \cite{WattsStrogatz} involves first creating a global structure in the form of a circular lattice of connections, before adding long-distance links. We aim to study a network growth model that depends on local interactions alone.

Consider the following model of how a person, {\bf $D$}, joining a social network might decide who she will becomes friends with first. Person $D$ comes in contact with one other person, $B$, chosen from all the people in the network at random. Rather than through a process of preferential attachment, person $B$ in this case is equally likely to be chosen if they have one friend as if they have two thousand. Person $D$ then asks person $B$ to recommend to her one other person, $C$ for her to become friends with. Again, person $B$ chooses $C$ at random from all the people they know, paying no attention to how many friends person $C$ has. In our model, person $D$ then becomes friends with either person $B$ (with probability $p$) or person $C$ (with probability $q$) and/or with both of them (with probability $p \cdot q$).

The motivation for analyzing this model is four-fold. Firstly, it is more parsimonious than the models we discussed above since it doesn't involve individuals counting contacts. It is truly local in terms of the decisions made by individual, i.e. at no point do they determine the degree of the people they attach to. Secondly, the model we describe is related to the friendship paradox \cite{Feld}. While $B$ and $C$ are both chosen at random, there is a subtle difference in how they are chosen which means that person $C$ is likely to have more friendships than person $B$. As we now show, and this is the third reason for analyzing this model, this subtle difference will prove sufficient to reproduce the three key properties: power-laws, clustering, and low degree of separation seen in real social networks. Finally, the fourth reason for studying this model is that when $p=0$ and $q=1$ (we only follow friends of friends), the model produces super-hub networks. The model thus generates, depending on the parameters, a whole range of different network structures.

 This paper is structured as follows. Before we make our own analysis of the 'friend of a friend' model, in the next section, we discuss the analysis of several related models. This is an important first step --- because while  other models have been proposed which are similar to or correspond to particular parameterisations of our model --- there are several prominent examples of misleading claims about how the model we discuss here can be used as a form of local mechanism for motivating preferential attachment. In the results section, we first simulate the model then derive a master equation for it, which allows us to explain power law behaviour and clustering in our model for $q=1$ and values of $p$ that are not too close to 1. We then look at the $p=0$ and $q=1$ case in particular, showing that it produces super-hub networks. We also prove that the distribution is non-stationary. Finally, we discuss our results in the context of fitting power laws to data, showing that our model produces power laws with empirical exponents ranging from $1.5$ upwards and clustering co-efficients ranging from 0 up to $0.5$. This is consistent with many real networks and we discuss how data from networks might be interpreted in light of our results.

 \section{Related models}
Models similar to, what we call, the friend of a friend model have been studied before in various contexts. The first very closely related model is the one proposed by Dorogovtsev and his colleagues, in which an edge of the graph is selected at random \cite {Dorogovtsev}.  They found that as time goes to infinity the probability that a randomly chosen node has degree $k$ is equal to
\[
\frac{12}{k(k+1)(k+2)}.
\]
i.e. the degree distribution has a power law of degree $3$. This is similar to the case where $p=1$ and $q=1$ in the friend of the friend model, but the distribution is different, since choosing one node at random gives a higher probability of choosing a node with lower degree than choosing an edge at random (as Dorogovtsev and his colleagues do).

Another related model is triadic closure, studied by for example Bianconi \textit{et al.} \cite {Bianconi}.  In this model, the first link is formed by connecting a new node to a randomly chosen existing target node and a second link is formed by {\it either} connecting the new node to a randomly chosen neighbour of the target with a probability $r$ {\it or} connecting to a completely different node with a probability $1-r$. When $r=1$ this model is equivalent to our model with $p=q=1$, but all other parameterisations differ from ours.

Jackson and Rogers \cite{Jackson} were the first researchers to study the friend of a friend model. They applied continuous time mean-field techniques to derive a master equation for the model of directed networks. They showed that when $p=1$ and $q=1$, the process gives a power-law of degree $3$.  Here we focus on the undirected case, but irrespective of this difference, as we shall soon see, Jackson and Rogers analysis is incomplete, and in some places incorrect, for the case where $p\ll1$. For example, they incorrectly predict that for $p=0$ and $q=1$ a power law with exponent $2$ will be obtained (see equation 5 and figure 1 in \cite{Jackson}). 

A similar mistake is made by Barab\'asi \cite{BAbook}, who claims in section 5.9 of his book Network Science, that for $p=0$ and $q=1$, this model is equivalent to the preferential attachment model and should thus give a power law with exponent $3$. The veracity of this last claim is important, because it is used as part of a discussion of what Barabasi calls 'the origins of preferential attachment'. The friend of a friend model is used by both Jackson and Rogers and Barabasi as an argument to get round the limitation, stressed in the first paragraph of the introduction of this paper, that preferential attachment is a global mechanism, while most social and biological processes are local. In short, there is thus a reasonably widespread misunderstanding of the potential outcomes of the friend of a friend model, which require clearing up.

One model, which Krapivsky and Redner call, 'growing networks with redirection' or GNR \cite{Krapivsky2001}, the friend to connect to is chosen uniformly at random with probability $1-r$ and the friend of that friend is chosen with probability $r$ \cite{Krapivsky2001}. $m$ edges are added for every new node.  This model is similar to our friend of the friend model with $p=1-r$ and $q=r$. There is, however, not a one to one mapping between our model and this type of model, since we do not have the constraint that $p+q=1$ and the average number of links added in our case is $m=p+q$. These authors appear to claim (in results section III) that this, specific model has an exponent of $1+\frac{1}{r}$ and thus, when $p=0$ and $q=1$, the exponent is predicted to be $2$. A similar result is also argued for a closely-related random walk model \cite {Vazquez2003}. This particular prediction is not, though, particularly central to Krapivsky and Redner \cite{Krapivsky2001}  article, which focussed mainly on what happens when $r=A_k$ is a linear function of $k$. 

 Krapivsky and Redner \cite{Krapivsky2005} went on to formulate a model where a new incoming node connects to a randomly selected target node, as well as to {\it all} out-going neighbours of the target node. Their algorithm is similar to a directed version of the one we will study, but instead of attaching to the uniformly random chosen friend-of-a-friend, it attaches to all nodes which that individual follows. In this case, the network generated by their model does give a power-law in-degree distribution with exponent $2$. 

In all of the above examples, the models either have a power-law stationary distribution, with an exponent which can be calculated using a master equation or mea-field approach \cite{Newmanbook2010} or, in cases where the underlying model does produce a non-stationary distribution, the authors do not analyse this property of the model.  Non-stationary networks often take the form of super-hubs, where one or a small number of nodes take all (in the sense of measure one) connections \cite {BAbook, Newmanbook2010}. This process is also referred to as condensation \cite{Bianconi2001, Dorogo2001,Dorogo2002}. As we outline in the introduction, such super-hub distributions are one property we will demonstrate in friend-of-a-friend networks. 

The first authors to explicitly state that models similar to friend-of-a-friend can produce non-stationary distributions, in the sense that there does not exist a limiting power-law (or any other stationary) distribution, were Gabel \textit{et al.} \cite {Gabel2013, Gabel2014}. In what they call an enhanced redirection model, a new node $u$ is either connected to a target node $v$ with the redirection probability $1-r$  or with $r$, node $u$ is connected to node $w$ that is followed by $v$. In this model, the parameter $r$ is a function of the degree of node $w$ and has the property that as  $r \to1$, the degree of node $w$ approaches infinity. The choice of $r \to1$ produces networks with several super hubs and non-stationary degree distributions with numerically estimated exponent values of $1<\alpha<2$. 

Similarly, Lambiotte \textit{et al.} \cite{Lambiotte} and Bhat \textit{et al.} \cite{Bhat}, independently of each other, implemented a copying process by introducing a probability $p_L$ of (undirected) link formation between each of the neighbours of a randomly selected target node and the new added node.  They observed sparse networks with power-law degree distributions for $p_L<\frac{1}{2}$. But, most interestingly, they found that for  $p_L\ge\frac{1}{2}$ the resulting degree distribution was non-stationary.  

After this, Krapivsky and Redner \cite{Krapivsky2017} studied the $p=0$ and $q =1$ case of the friend of a friend model we look at here. The key features of this model is that it generates non-stationary networks with  super-hub nodes. They found that the degree distribution has an exponent $\alpha$ that is strictly less than 2, i.e. a power law with no mean. Specifically, their analysis show that the number of nodes with degree $k$, $N_k$ scales as 
    \[ N_k \sim N^{\alpha-1}, \quad  \text{for} \quad k\gg2 \] 
  and \[ N_k \sim N^{\alpha-1}k^{-\alpha}, \quad  \text{for} \quad k\gg1, \]
  where $N$ is the network size. The value of $\alpha$ was approximated numerically and found to be $1.566$. 
  
Taking the literature mentioned above, together with further variations on this theme, we find a rich variety of useful results about models closely related to the 'friend of a friend' model we will now investigate. Most importantly, Krapivsky and Redner \cite{Krapivsky2017} have already provided useful numerical results about the case where $p=0$ and $q=1$. 

It is thus important that we are clear about what we see as the contribution of the current work. We see the 'friend of a friend' idea as having a special role, because the mechanism is both local (it requires no information about the degree of any of the nodes attached to) and very simple to state (simpler even than the original network models \cite{BA, WattsStrogatz,Holme}). It has also been used to motivate the origins of these more complicated/non-local models. Our contribution then is thus to provide a self-contained description of the important properties of the 'friend of a friend' model – which are surprisingly varied – and a mathematical proof of the non-stationary nature of the distribution when $p=0$ and $q=1$.

\section{Results}

\section*{Simulation model}
\begin{figure}
	\centering
	\vspace*{-50mm} 
	\hspace*{-25mm} 
	\includegraphics[width=\textwidth]{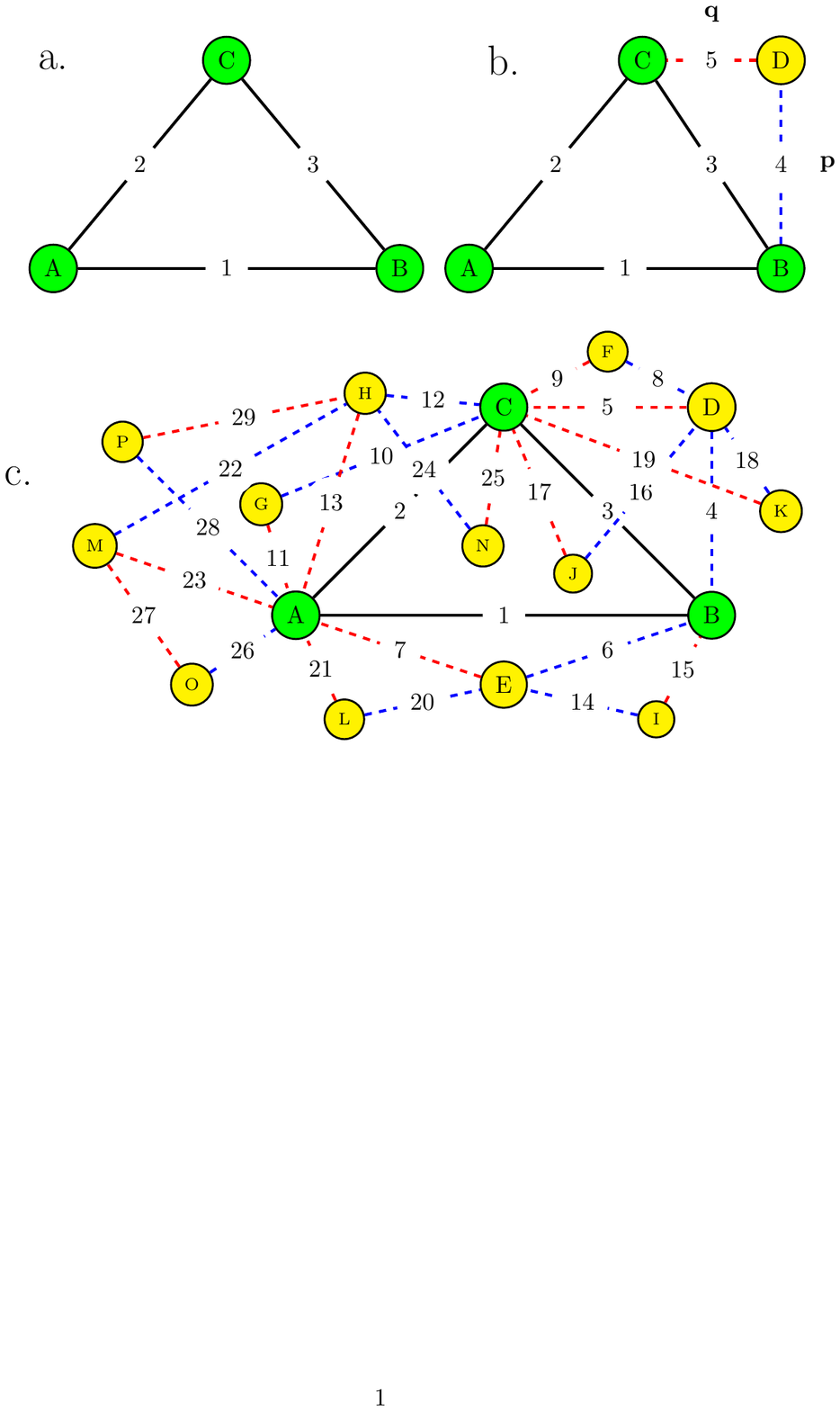}
	\vspace*{-105mm} 
	\caption{\textbf{Schematic illustration of the formation
				and evolution of the network}. Initial nodes are represented by green shading and new incoming nodes are shown by yellow shading. Black solid lines are initial links, blue dashed lines are the new links formed with probability $p$ (picking a friend) and red dashed lines are the new links created with probability $q$ (picking a friend of a friend). Nodes are labelled alphabetically and linked with edges labelled in numbers.  As shown in Fig. \ref{Network formation}a, network formation starts at $t=3$. Fig. \ref{Network formation}b shows that when the 4th node (labelled $D$) is added it first picks node $B$ as a friend (blue dashed line of link number 4) then it picks node $C$ as a friend of node $B$ (red dotted line of link number 5).  Fig. \ref{Network formation}c illustrates how this formation continues in which new incoming node join two earlier nodes with $p = 1$ and $q = 1$. So, node $E$ links to two old nodes by first picking node $B$ as a friend and then $A$ as a friend of $B$, node $F$ make link $8$ with $D$ then link $9$ with node $C$, etc.  The formation continues until the last node $P$ is connected to node $A$ as a friend and $H$ as a friend of $A$}
	\label{Network formation}
\end{figure}

We now formalise the model described in the introduction. The mechanism of formation of the network is illustrated in Fig. \ref{Network formation}. 
We start with a network of three mutually connected friends, as  shown in Fig. \ref{Network formation}a. We add one friend (or node) to the network per time step, using $t$ to denote both the time step and the number of individuals. The new node ($D$) is added as follows. First it picks one of the existing nodes ($B$) uniformly at random from all the $t$ nodes as shown in Fig. \ref{Network formation}b. The node $D$ then forms an attachment to $B$ with probability $p$. After that it picks randomly a neighbour ($C$) of the original node ($B$) and attaches to it with probability $q$. In the simplest version of this model, $p=q=1$, i.e, $D$ attaches to both $B$ and $C$. Fig. \ref{Network formation}c shows a typical network example after $t=16$ time steps for the same extreme situation $p=q=1$. 

\begin{figure}
	\includegraphics[width=\textwidth]{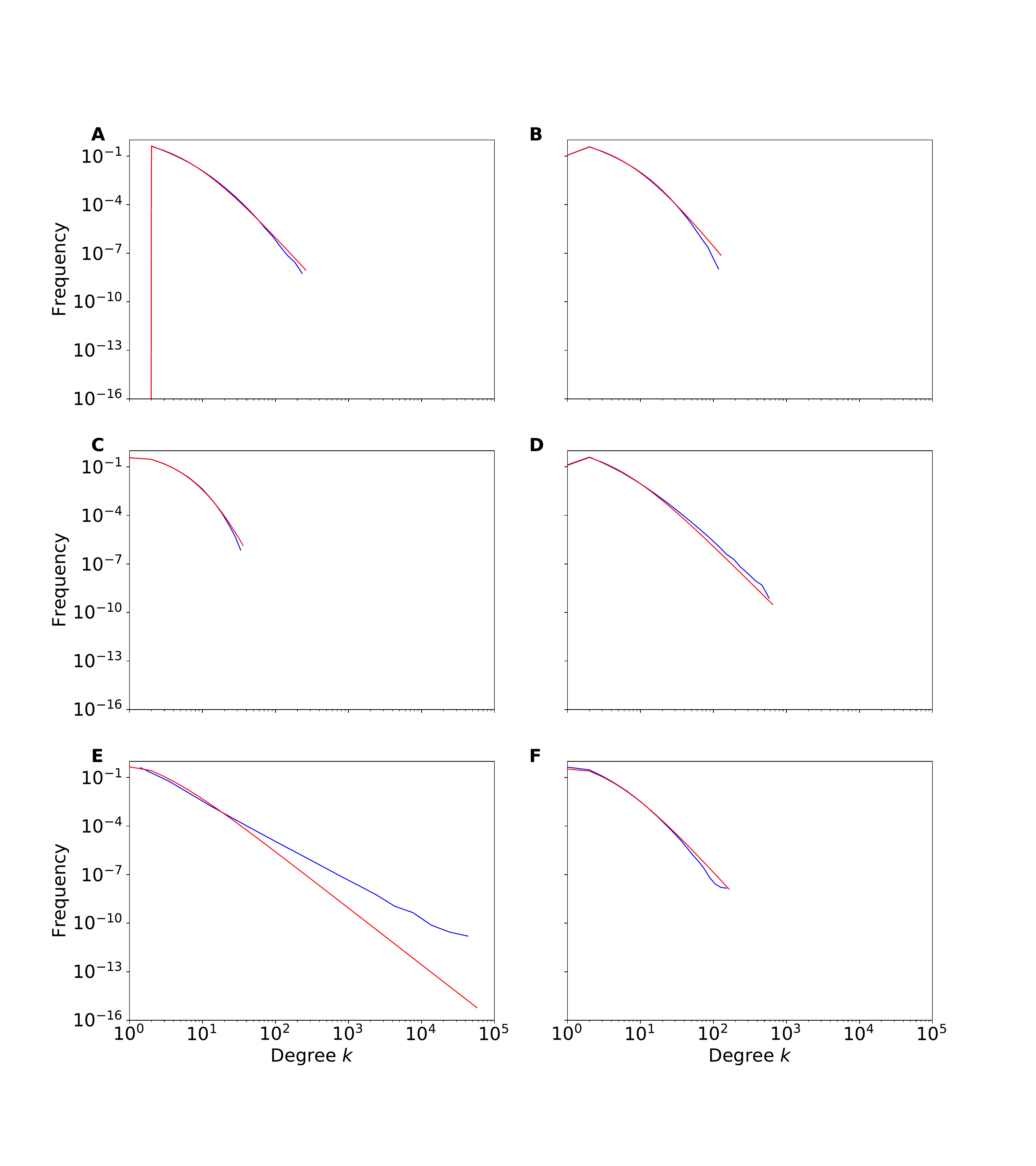}
	\vspace*{-35mm}
	\caption{\textbf{Numerical and theoretical degree distributions}. The blue lines are the numerical simulation results and the red solid lines  represent predicted values. The numerical simulations are averaged over 5 runs for each network with $t=10^{6}$ nodes. The parameter values used in each Figure are A : $p = 1$, $q = 1$, B : $p = 1$, $q =0.75 $, C : $p = 1$, $q = 0.25$, D : $p = 0.75$, $q = 1$, E : $p = 0.25$, $q = 1$ and F : $p = 0.5$, $q = 0.5$}
	\label{Degree distributions}
\end{figure}

In order to get an overall impression of the degree distributions generated by this model, we first run the simulation for representative parameter values and looked at the resulting degree distribution. We use logarithmic binning to plot the histograms on logarithmic scales using the method of Newman \cite{Newman}.  The outcome of the simulations are shown as the blue lines in Fig. \ref{Degree distributions}.

We see that for $p=1$ and $q=1$ (Fig. \ref{Degree distributions}A), where we attach to both the randomly chosen person and the randomly chosen friend, for degree $k>10$, we get a relatively straight line in the log-log plot of the frequency distribution. As we decrease $q$, while keeping $p=1$ constant, the frequency of nodes with higher degree decreases (Figures \ref{Degree distributions}B and \ref{Degree distributions}C). Conversely, if we decrease $p$, while keeping $q=1$ constant, higher degree nodes become more frequent (Figures \ref{Degree distributions}D and \ref{Degree distributions}E). For $p=q=0.5$ some nodes are not connected, but the ones that are have a frequency distribution reasonably similar to $p=q=1$.

\section*{Master equation}

In order to better understand how the degree distribution depends on the parameters $p$ and $q$, we create a master equation estimate of the proportion of individuals with $k$ friendships at integer time $t$ as $P_k(t)$ \cite{Newman}. We treat $t=3$ as the first time step at which $P_2(3)=3$ and $P_k(3)=0$ for all other values of $k$ (as in Fig. \ref{Network formation}a). After $t>3$ time steps there will be $t$ nodes and on average $t(p+q)-3$ edges. In the limit of $t \to \infty$ the expected ratio of edges to nodes tends to $(p+q)$. 

We now consider the probability distribution for the number of neighbours of the chosen node $B$ and $C$ in turn. Since $B$ is chosen uniformly at random, the probability that it has $k$ neighbours is simply $P_k(t)$, i.e. the proportion of nodes with $k$ neighbours. Now, let $K_C$ denote the number of neighbours of node $C$, the neighbours of $B$ which is chosen. We want to know $\mbox{P}(K_C=k | \mbox{$B$ friends with $C$})$, i.e.  the probability that $C$ has $k$ neighbours given that it is friends with $B$. This can be obtained using Bayes' Theorem,
\begin{equation}
\mbox{P}(K_C=k | \mbox{$B$ friends with $C$}) =  \frac{\mbox{P}( \mbox{$B$ friends with $C$} | K_C=k ) \cdot P(K_C=k)}{\sum_j \mbox{P}( \mbox{$B$ friends with $C$} | K_C=j ) \cdot \mbox{P}(K_C=j)} 
\end{equation}
If we assume that $B$ is equally likely to be friends with any person in the network (we will come back to this assumption later, which does not always hold), then the probability $B$ friends with $C$ is  
\begin{equation}
\mbox{P}( \mbox{$B$ friends with $C$} | K_C=k ) = \frac{k}{t} 
\end{equation}
since $C$ friends with $k$ out of all the $t$ individuals in the network. It thus follows that
\begin{eqnarray*}
\mbox{P}( K_C =  k | \mbox{$B$ friends with $C$}) & = & \frac{(k/t) \cdot \mbox{P}(K_C=k)}{\sum_j (j/t) \cdot \mbox{P}(K_C=j)} \\
& = & \frac{k \cdot \mbox{P}(K_C=k)}{\mbox{E}[K_C]} \\
& = & \frac{k \cdot P_k(t)}{\mbox{E}[K_C]} 
\end{eqnarray*}
where $P_k(t)$ is the prior probability of a node having $k$ neighbours.  We can calculate $\mbox{E}[K_C]$ by noting that the mean or expected number of degrees added each time step is $2(p+q)$ (since edges are bidirectional). Thus, finally,
\begin{equation}
\mbox{P}( K_C=k | \mbox{$B$ friends with $C$}) = \frac{k \cdot P_k(t)}{2(p+q)}  \label{BayesBC}
\end{equation}
gives the probability of attaching to a node with degree $k$.

The average number of attachments to a node with degree $k$  is thus
\begin{equation}  \label{Equation0a11q}
 p P_k(t) + q \frac{k \cdot P_k(t)}{2(p+q)} 
\end{equation} 
where the first term is attachment to the randomly chosen node, $B$, and the second term is attachment to the neighbour of $B$ , the node $C$. Following Newman \cite{Newman}, we can now write a master equation for the evolution of the proportion of individual with $k \geq 3$ friendships as
\begin{equation}  \label{Equation0a11}
	(t +1)P_{k}(t + 1) =  tP_k(t) + (p+\frac{q}{2(p+q)} (k-1))P_{k-1}(t)- (p+\frac{q}{2(p+q)}k)P_k(t) , \quad \text{for} \quad k\ge 3.
\end{equation}
The first term on the right hand side ($tP_k(t)$) gives the number of nodes with degree $k$. The second term describes the outcome when a node with degree $k-1$ is linked to, and the third term is the outcome when a node with degree $k$ is linked to. 

Since edge formation between the new incoming node and the existing ones depend on $p$ and $q$, it is also possible for the growing network $N(k,t)$ to have nodes with $k<3$.  We can deduce the equations for the proportion of nodes with $k=0$, $k=1$ and $k=2$ as follows. For $k=0$, isolated nodes appear in the network when neither the target nor the neighbour node is chosen, i.e. with probability $(1-p)(1-q)$, the network may be disconnected and may consist
of several isolated components. At each time step the proportion of isolated nodes $P_0(t)$ will increase on average by $(1-q)(1-p)$ and decrease with $pP_0(t)$, giving 
 \begin{equation}  \label{Equation0a11a}
 	(t +1)P_{0}(t + 1) =  tP_0(t) + (1-q)(1-p)- pP_0(t), \quad \text{for} \quad k=0.
 \end{equation}

There are three possible cases for which a new incoming node chooses the target node and creates only one edge. The first case occurs with probability $p(1-q)$, i.e. when a target node is chosen but the neighbour is not. The second case, occurs with probability $(1-p)q$, i.e. when a target node is not chosen but the neighbour is. As a result, the proportional of nodes with $k=1$ is increased on average by $p(1-q)+q(1-p)$. The third case is when the incoming node chooses an existing isolated ($k=0$) target node and this occurs with probability $pP_{0}(t)$. The master equation of such nodes with degree $k=1$ is given by 
\begin{equation}  \label{Equation2bA11}
	(t +1)P_{1}(t + 1)	=  tP_1(t) + p(1-q)+q(1-p) +pP_{0}(t) - (p+\frac{q}{2(p+q)})P_1(t) , \quad \text{for} \quad k=1.
\end{equation}
nodes with degree $k=2$ appear when the new incoming node makes connection with both the target and the neighbour, i.e. with probability  $p\cdot q$. Otherwise, the master equation is similar to that for $k \geq 3$, i.e.
\begin{equation}  \label{Equation0a11a2}
	(t +1)P_{2}(t + 1) =  tP_2(t) + pq+(p+\frac{q}{2(p+q)})P_{1}(t)- (p+\frac{q}{(p+q)})P_2(t) , \quad \text{for} \quad k=2.
\end{equation}

Taking the limit $t\to \infty$ so that $P_{k}(\infty) \to P_{k}$, Eqs. (\ref{Equation0a11}), (\ref{Equation0a11a}),  (\ref{Equation2bA11}) and (\ref{Equation0a11a2}) simplify to
\begin{equation} \label{combination}
	P_{k} =
	\begin{cases}
		\frac{(1-q)(1-p)}{1+p}, & \quad \text{for} \quad k=0, \\
		\frac{p(1-q)+q(1-p)+pP_{0}}{1+p+\frac{q}{2(p+q)}}, & \quad \text{for} \quad k=1, \\
		 \frac{pq+ (p+\frac{q}{2(p+q)})P_{1}}{1+p+\frac{q}{p+q}}, &\quad \text{for} \quad k=2, \\
		\frac{k-1 +\frac{2p(p+q)}{q}}{k+ \frac{2(p+q)}{q}+\frac{2p(p+q)}{q}}P_{k-1}, & \quad \text{for} \quad k\ge3.\\
	\end{cases}       
\end{equation}

Expanding Eq. (\ref{combination}) we obtain a general asymptotic degree distribution,
\begin{equation}  \label{Equation4bA11}
	\begin{split}
	P_k &= \frac{(k-1+\frac{2p(p+q)}{q})(k-2+\frac{2p(p+q)}{q})...\times (2+\frac{2p(p+q)}{q})}{(k+\frac{2(p+q)}{q}+\frac{2p(p+q)}{q})(k-1+ \frac{2(p+q)}{q}+\frac{2p(p+q)}{q})...\times(3+\frac{2(p+q)}{q}+\frac{2p(p+q)}{q}) } \times P_{2} \\
	\end{split}
\end{equation}
where $P_2$ is a constant value which depend on the values of $p$ and $q$. It is computed from
\begin{equation}  \label{p2}
P_2 = \frac{(1+\frac{2p(p+q)}{q})(\frac{2p(p+q)}{q})\left[\frac{q(1+p) \left[q+2(p+q)(1+p)\right] }{q+2p(p+q)}+ 2(1-q)+\frac{q(1-p^2)}{p} \right]}{(2+\frac{2(p+q)}{q}+\frac{2p(p+q)}{q})(1+ \frac{2(p+q)}{q}+\frac{2p(p+q)}{q})(1+p) }
\end{equation}
Using the gamma function, we write Eq. (\ref{Equation4bA11})  as follows, 
\begin{equation}  \label{Equation5bA11}
	P_k = P_2 \frac{\Gamma(k+\frac{2p(p+q)}{q})\Gamma(3+\frac{2(p+q)}{q}+\frac{2p(p+q)}{q})}{\Gamma(2+\frac{2p(p+q)}{q})\Gamma(k+1+ \frac{2(p+q)}{q}+\frac{2p(p+q)}{q})}.
\end{equation}
Multiplying $\Gamma(\frac{2(p+q)}{q}+1)$ on both the numerator and denominator of Eq. (\ref{Equation5bA11}) to obtain,
\begin{equation}  \label{Equation6bA1}
	P_k= P_2 \frac{\Gamma(k+\frac{2p(p+q)}{q})\Gamma(1+ \frac{2(p+q)}{q})}{\Gamma(k+1+\frac{2(p+q)}{q} + \frac{2p(p+q)}{q})} \times \frac{\Gamma(3+\frac{2(p+q)}{q}+\frac{2p(p+q)}{q})}{\Gamma(2+\frac{2p(p+q)}{q})\Gamma(1+\frac{2(p+q)}{q})}\\
\end{equation}
Using the beta function, we write Eq. (\ref{Equation6bA1})  as follows,
\begin{equation}  \label{betaequation}
P_k=P_2\frac{\Gamma(3+\frac{2(p+q)}{q}+\frac{2p(p+q)}{q})}{\Gamma(2+\frac{2p(p+q)}{q})\Gamma(1+\frac{2(p+q)}{q})} B(k+\frac{2p(p+q)}{q},1+\frac{2(p+q)}{q}).
\end{equation}
The beta function has the property that when $x\to\infty$ and $x\gg y$ then
$B(x,y)\approx $ $x^{-y}\Gamma(y)$. Thus, when $k$ is very large, $P_k$ can be approximated as 
\begin{equation}  \label{betaequation1}
P_k \approx  P_2\frac{\Gamma(3+\frac{2(p+q)}{q}+\frac{2p(p+q)}{q})}{\Gamma(2+\frac{2p(p+q)}{q})} (k+\frac{2p(p+q)}{q})^{-(1+\frac{2(p+q)}{q})}.
\end{equation}
When the values of $\frac{2(p+q)}{q}$ and $\frac{2p(p+q)}{q}$ are integers we have,

\begin{equation}  \label{gammaequation}
	P_k \approx  P_2(2+\frac{2(p+q)}{q}+\frac{2p(p+q)}{q})(1+\frac{2(p+q)}{q}+\frac{2p(p+q)}{q})...(2+\frac{2p(p+q)}{q}) (k+\frac{2p(p+q)}{q})^{-(3+\frac{2p}{q})}.
\end{equation}
\begin{equation} \label{powerA11}
	P_k  \sim  k^{-\alpha}
\end{equation} 
where $\alpha = 3+\frac{2p}{q}\ge3$. In Appendix A we derive this same result in another way, using the mean-field approach.

When $p=q=1$ the solution of Eq. (\ref{combination}) for the case $k\ge3$ becomes exact given by, 
\begin{equation}  \label{Exacts}
\begin{split}
P_k &= \frac{(k+3)(k+2)...\times 8\times 7\times 6}{(k+8)(k+7)...\times 13\times 12\times11 } \times P_{2} \\
&=\frac{12096}{(k+8)(k+7)(k+6)(k+5)(k+4)} 
\sim  k^{-5}
\end{split}
\end{equation}

The red lines in Fig. \ref{Degree distributions} show the distribution derived from the above master equation. For $p=1$ and $q=1$ we find a similar degree distribution as in the simulation, with a power-law exponent $\alpha = 5$ (Fig. \ref{Degree distributions}A). However, the numerical estimate of the exponent of the power-law ($\alpha=5$) is not obtained for $p=q=1$ because the maximum degree is around $k=100$, which is not sufficiently large to reliably measure a power-law.

Decreasing $q$ to $0.75$ and $0.25$ while holding $p=1$ (Figures \ref{Degree distributions}A and \ref{Degree distributions}B respectively), we again see similar trend in the analytical results as those observed in numerical calculations, although the slope of the distribution in the simulation is slightly steeper.  However, Figures \ref{Degree distributions}D and, especially, \ref{Degree distributions}E show that $q$ when is maintained at $1$ and $p$ is reduced to $0.75$ and $0.25$  the slope of the distribution is less steep in the simulation than in the theory (we will return to this point in a later section). In Fig. \ref{Degree distributions}F where $p=q=0.5$, we still have $\alpha = 5$ as that of Fig. \ref{Degree distributions}A. We can also see that the slope of the numerical line is reasonably close to the analytical one.

\subsection*{Clustering}
The local clustering coefficient ($C_v$) of an arbitrary node $v$ with degree $k_v$ is measured as the ratio between the number of all edges that exist among all first neighbours of $v$ ($E_v$) and the maximum number of potential links between the neighbours of $v$ \cite {WattsStrogatz}, which is $k_v(k_v-1)/2$. Thus, 

\begin{equation}  \label{Equation1}
C_v =  \frac{2E_v}{k_v(k_v-1)} \; .
\end{equation}

We are interested in how the average local clustering coefficient depends on degree $k$. Hence we define a mean clustering over all nodes with degree $k$:
\begin{equation}
C_k = \frac{2E_k}{k(k-1)} \; ,
\end{equation}
where
\begin{equation}
E_k = \frac{\sum_{v:k_v=k} E_v}{t P_k} \; .
\end{equation}

\begin{figure}
	\centering
	\vspace*{-28mm} 
	\includegraphics[width=\textwidth]{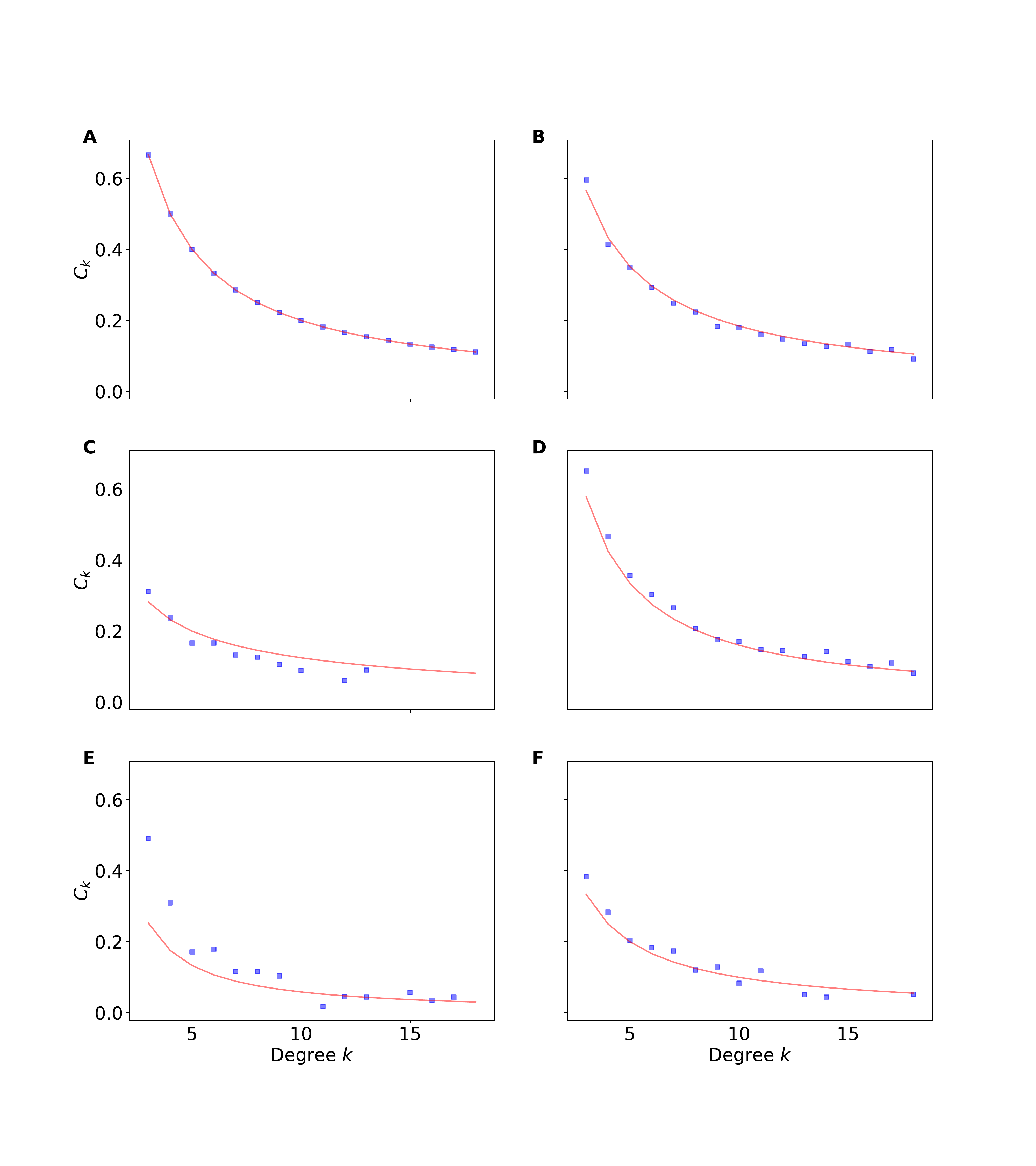}
	\vspace*{-35mm} 
	\caption{\textbf{Comparisons of numerical simulations and theoretical calculations for clustering coefficients as function of node's degree $k$.} In each panel, we use the following parameters. A : $p = 1$, $q = 1$, B : $p = 1$, $q =0.75 $, C : $p = 1$, $q = 0.25$, D : $p = 0.75$, $q = 1$, E : $p = 0.25$, $q = 1$, F : $p = 0.5$, $q = 0.5$}
	\label{Clustering_model_k}
\end{figure}

The blue points in Fig. \ref{Clustering_model_k} show how the mean local clustering coefficient changes with degree $k$ for different attachment probabilities, $p$ and $q$. In all cases, the local clustering decreases as degree increases. When $p=1$ and $q=1$ (Fig. \ref{Clustering_model_k}A) the clustering decays more slowly than when $p<1$ or $q<1$ (Figures \ref{Clustering_model_k}B-\ref{Clustering_model_k}F). When $p=0.25$ and $q =1$, or when $p=q=0.5$ there is much less clustering, especially for nodes with high degree (Figures \ref{Clustering_model_k}E and \ref{Clustering_model_k}F).

We now derive an analytical approximation of the local clustering coefficient based on a method similar to the master equation. We consider the attachment illustrated in Fig. \ref{Network formation}b, where the network has started growing. We first look at the case where $p=q =1$ and will then adapt this result to the more general case. Whenever a new node is attached to an existing $v$, the degree of $v$ increases by $1$ and $E_v$ also increases by $1$, because we also attach to one of $v$ neighbours (assuming $p=q=1$). Furthermore, all nodes have an initial value of $E_v=1$ and $k_v=2$ (since the two nodes they attach to are attached to each other with one edge, as in Fig. \ref{Network formation}a). Thus,
 \begin{equation}  \label{clusterEquatione}
 E_k= k-1
 \end{equation}
Therefore,
\begin{equation}  \label{clusterEquationa}
C_k= \frac{2}{k}.
\end{equation}
This equation matches the simulation results, suggesting that the approximation is accurate (red line in Fig. \ref{Clustering_model_k}A).
 
When either $p<1$ or $q<1$, there are two ways in which a node $v$ can increase the total number of links between its neighbours. The first is when $v$ is the first node chosen (at random from all nodes), and then a connection is made to both $v$ and its chosen neighbour. This occurs with probability
\[p P_k(t)q.
\] 
The second way is when the first node is a neighbour of $v$, from which $v$ is then chosen as the second potential attachment. This occurs with probability
\[ 
q \frac{k \cdot P_k(t)}{2(p+q)}p.
\]
Summing these two possible cases gives,
\begin{equation}  \label{clusterEquation1}
p P_k(t)q + q \frac{k \cdot P_k(t)}{2(p+q)}p
\end{equation} 
which is the expected average number of triangles attached to a node with degree $k$. The ratio between this (Eq. (\ref{clusterEquation1})) and the overall average number of attachments to the node (Eq. (\ref{Equation0a11q})) gives the expected increase in $E_v$ each time an edge is added to a node of degree $k$. As a result,
\begin{equation}  \label{clusterEquation3}
E_k = \frac{p P_k(t)q + q \frac{k \cdot P_k(t)}{2(p+q)}p}{ p P_k(t) + q \frac{k \cdot P_k(t)}{2(p+q)} } \cdot (k-1)  = \frac{2(p+q)+k}{\frac{2(p+q)}{q}+\frac{k}{p}}\cdot (k-1) \quad \text{for} \quad k \ge3.
\end{equation}
Thus, Eq. (\ref{clusterEquationa}) now becomes,
\begin{equation}  \label{clusterEquation3a}
C_k= \frac{2(p+q)+k}{\frac{2(p+q)}{q}+\frac{k}{p}}\cdot \frac{2}{k} = \frac{4(p+q)+2k}{\frac{2k(p+q)}{q}+\frac{k^2}{p}}  \quad \text{for} \quad k \ge3.
\end{equation}
The prediction based on this approximation is shown as the red line in Fig. \ref{Clustering_model_k}B-\ref{Clustering_model_k}F and partially capture the numerical results (given by blue squares).

\begin{figure}
	\centering
		\vspace*{-8mm} 
	\includegraphics[width=\textwidth]{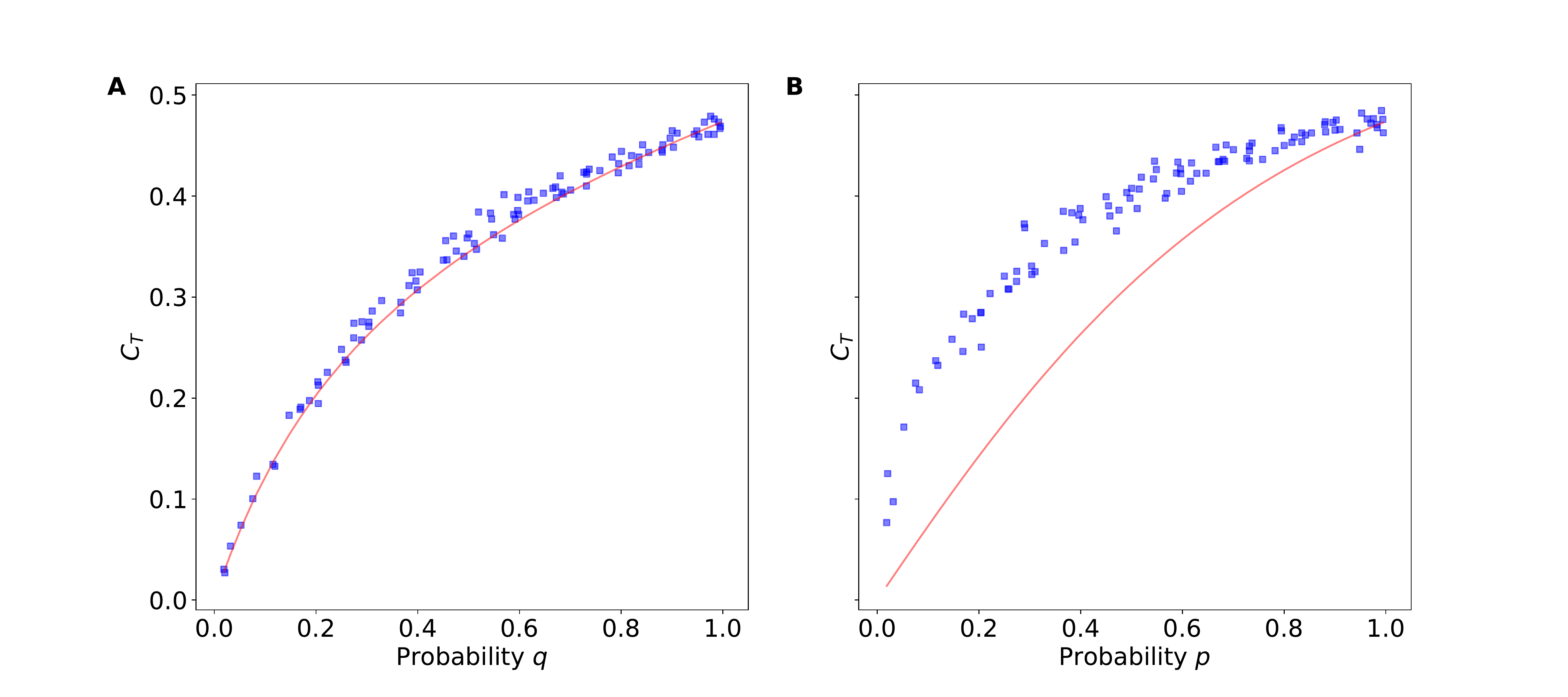}
	\vspace*{-20mm} 
	\caption{ \textbf{Comparisons of numerical simulations and theoretical calculations for clustering coefficient computed as a function of probabilities.} Parameters used in panel $A$ are $p=1$ and $0 \le q \le 1$ while those of panel $B$ are $0 \le p \le 1$ and $q=1$}
	\label{Clustering_model3}
\end{figure}

The network clustering coefficient $C_T$ is the average of local clustering coefficients $C_k$ given that $k\geq 3$. This can be written as, 
\begin{equation}  \label{Equation4b} 
C_T = \frac{1}{1-(P_2+P_1+P_0)}\sum\limits_{k=3}^{\infty}P_kC_k.
\end{equation}
When $p=q=1$ we have,
\[
C_T = \frac{1}{1-2/5}\sum\limits_{k=3}^{\infty} \frac{2}{k}\frac{12096}{(k+8)(k+7)(k+6)(k+5)(k+4)}  = \frac{199}{420}   \approx 0.4738
\]
Fig. \ref{Clustering_model_k}A compares the numerical simulation for the model to this sum for $p=1$ and varying $q$. We see a good agreement between the theoretical predictions and the numerical results. Holding $q=1$ constant and varying $p$, we see a significant deviation of the theoretical values from numerical ones (Fig. \ref{Clustering_model3}B) with the only intersection at around $p =1$. 
 
\subsection*{Super-hub networks}

While the master equation method gives a good approximation to the degree distribution and clustering for a range of values of $p$ and $q$, it fails when $q$ is close to one and $p$ is close to zero (for example, Fig. \ref{Degree distributions}E). This is at first surprising, because in, for example, Barab\'asi's book {\em Network Science}, there is a claim that this model with $p=0$ and $q=1$ should produce a power-law with degree $3$ ( see 'related models' for details) \cite {BAbook}. Similar claims can be found in, for example, \cite{ Jackson}. Indeed, Barab\'asi refers to the friend of a friend model, which he calls the copying model, as providing the origins of preferential attachment. Why then does the simulation not agree with the master equation?

	\begin{figure}
		\centering
		\begin{subfigure}[b]{0.49\textwidth}
			\centering  
			\subcaption[font=large]{}		
			\includegraphics[width=\textwidth]{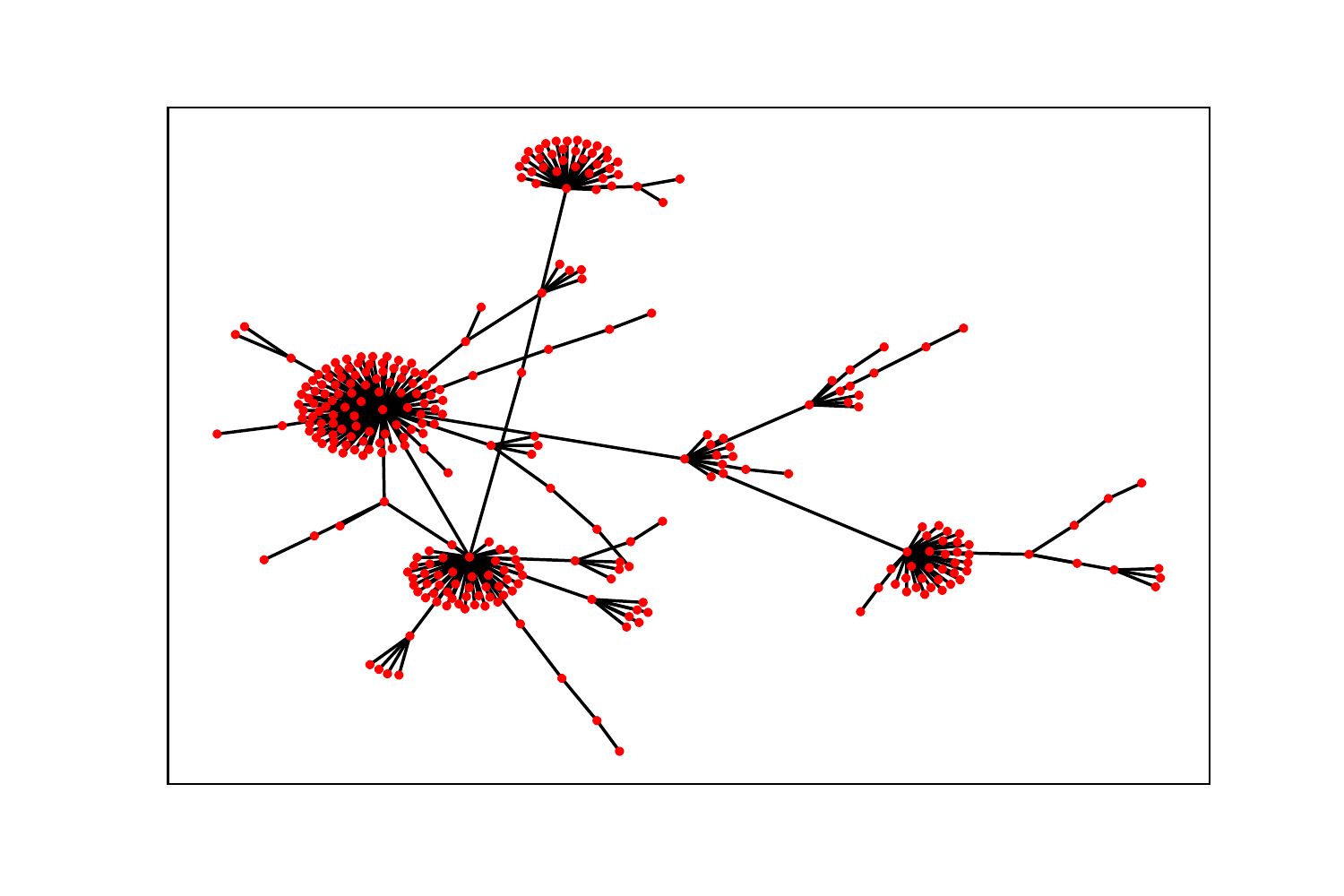}    			    
			\label{fig:N1}
		\end{subfigure}
		\hfill
		\begin{subfigure}[b]{0.49\textwidth}
			\centering
			\subcaption[textfont=large]{}
			\includegraphics[width=\textwidth]{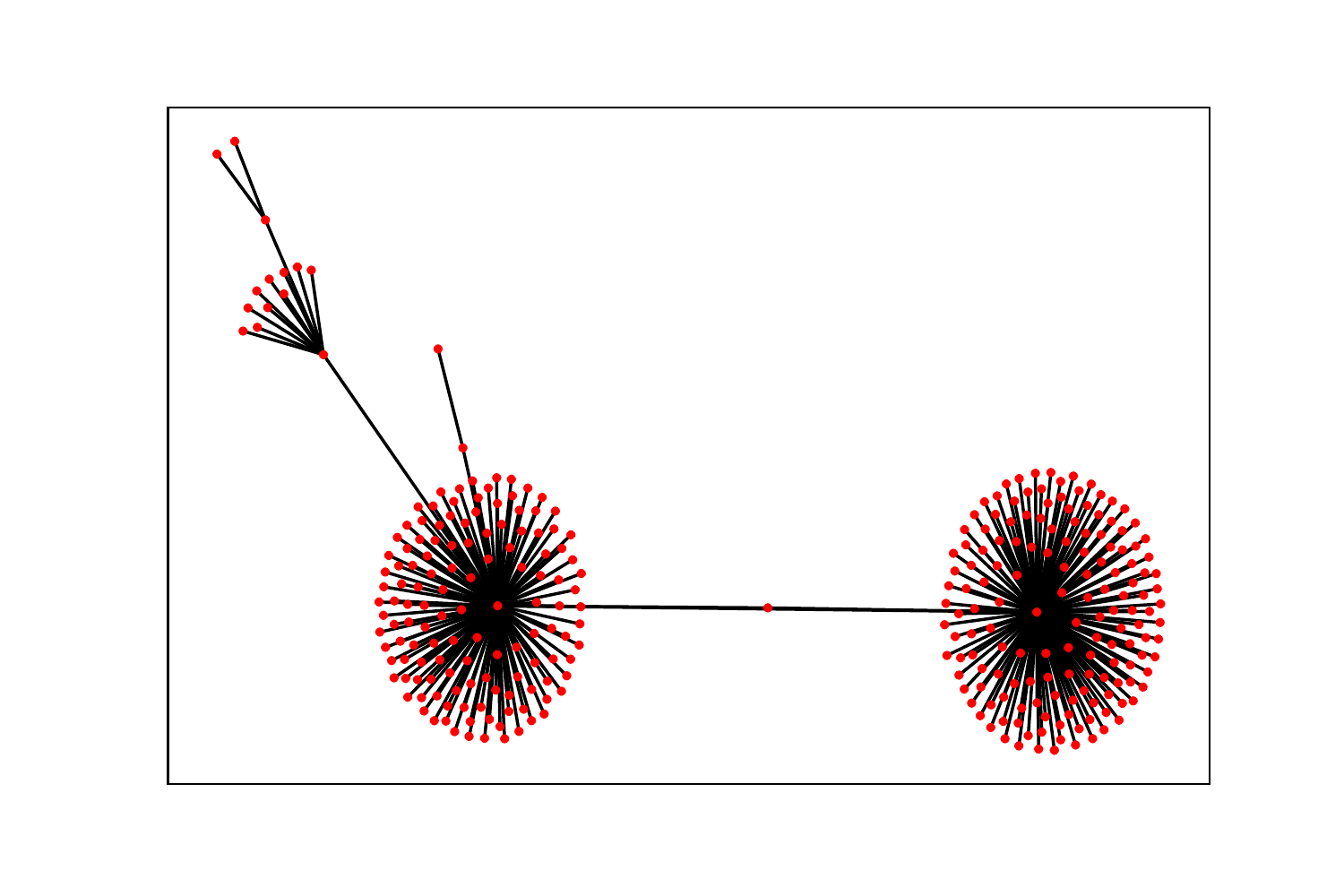}
			\label{fig:N2}
		\end{subfigure}
		\vfill
		\begin{subfigure}[b]{0.49\textwidth}
			\centering
			\subcaption[textfont=large]{}
			\includegraphics[width=\textwidth]{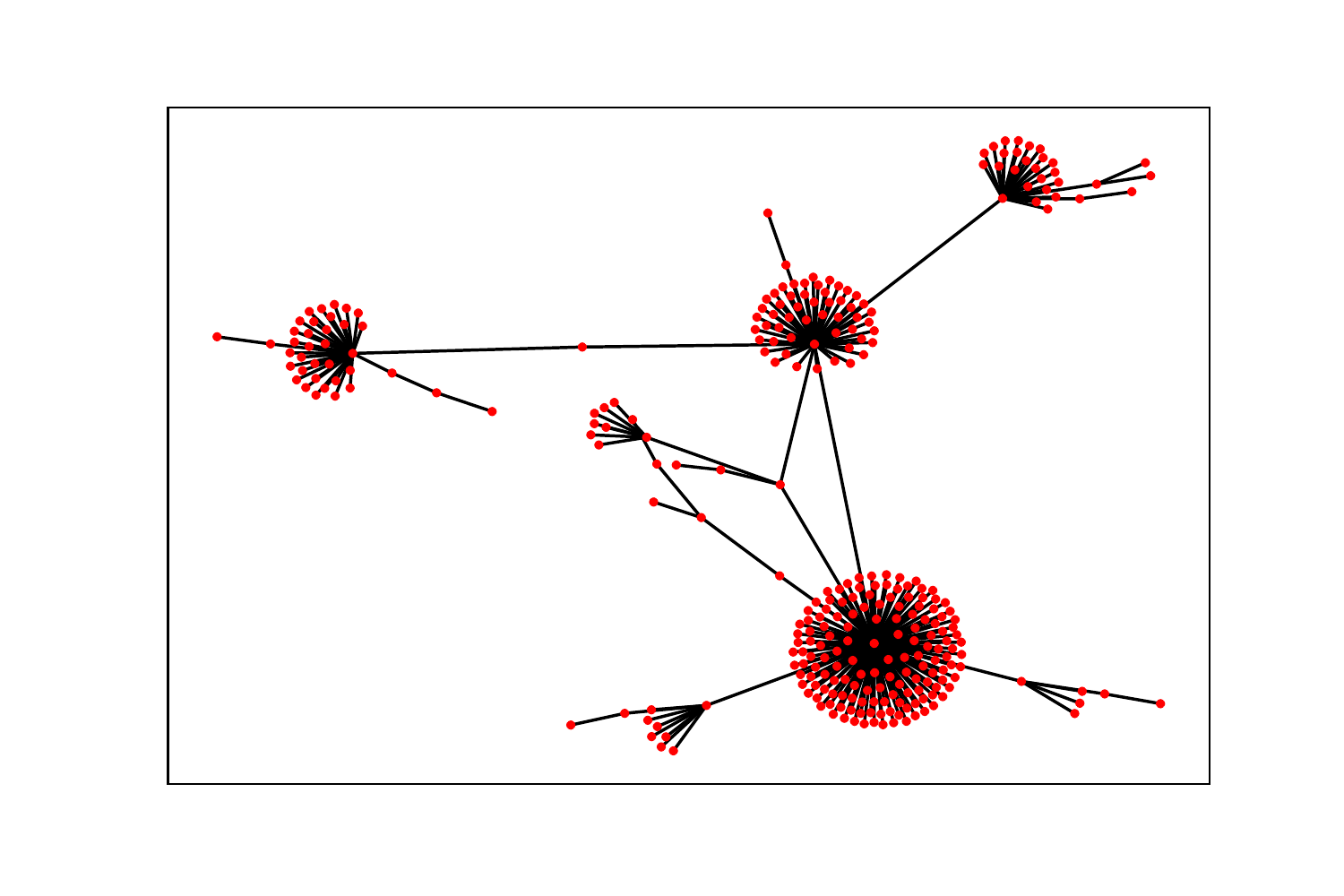}
			\label{fig:N3}
		\end{subfigure}
		\hfill
		\begin{subfigure}[b]{0.49\textwidth}
			\centering
			\subcaption[textfont=large]{}
			\includegraphics[width=\textwidth]{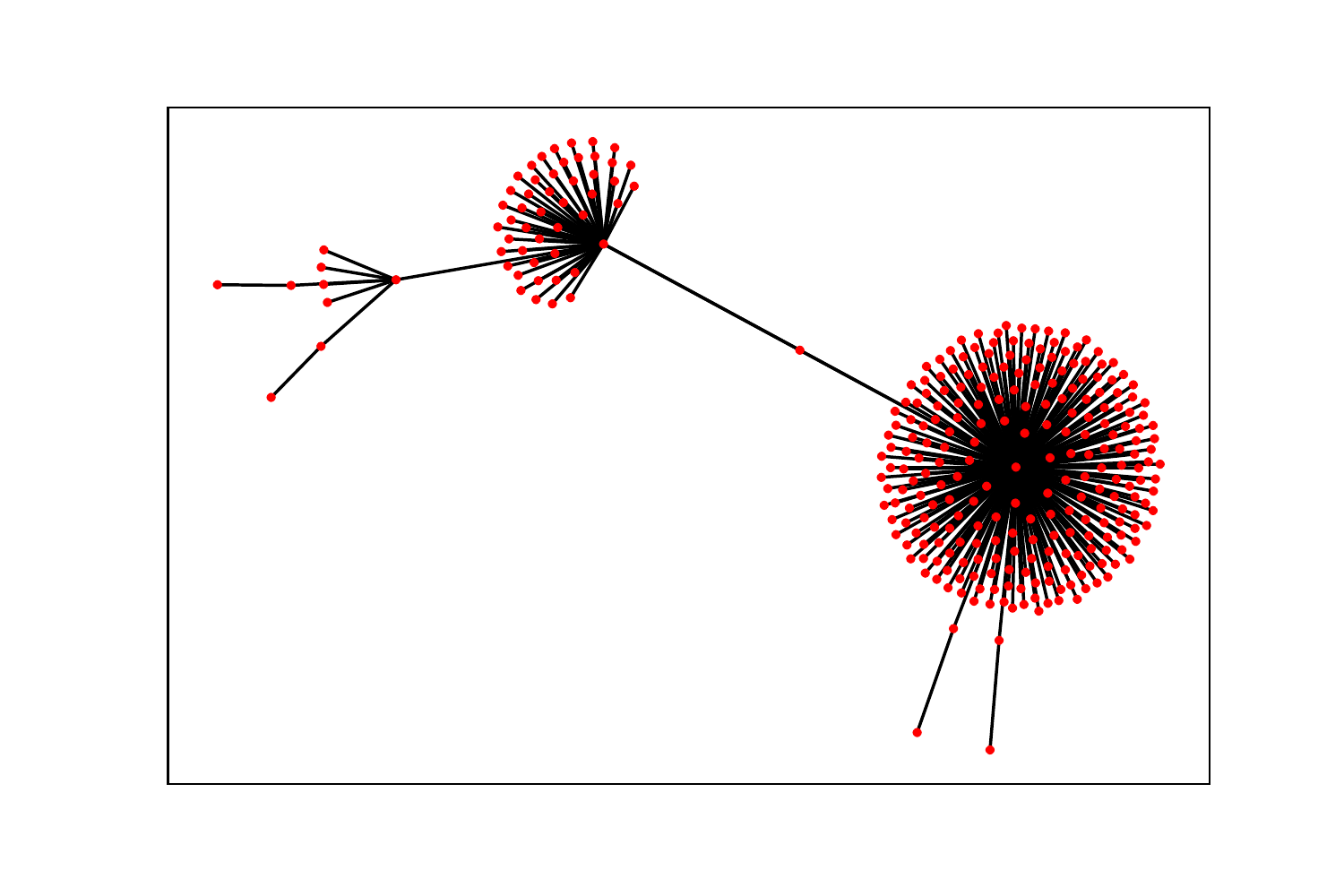}
			\label{fig:N4}
		\end{subfigure}
		\vfill
		\begin{subfigure}[b]{0.49\textwidth}
			\centering
			\subcaption[textfont=large]{}
			\includegraphics[width=\textwidth]{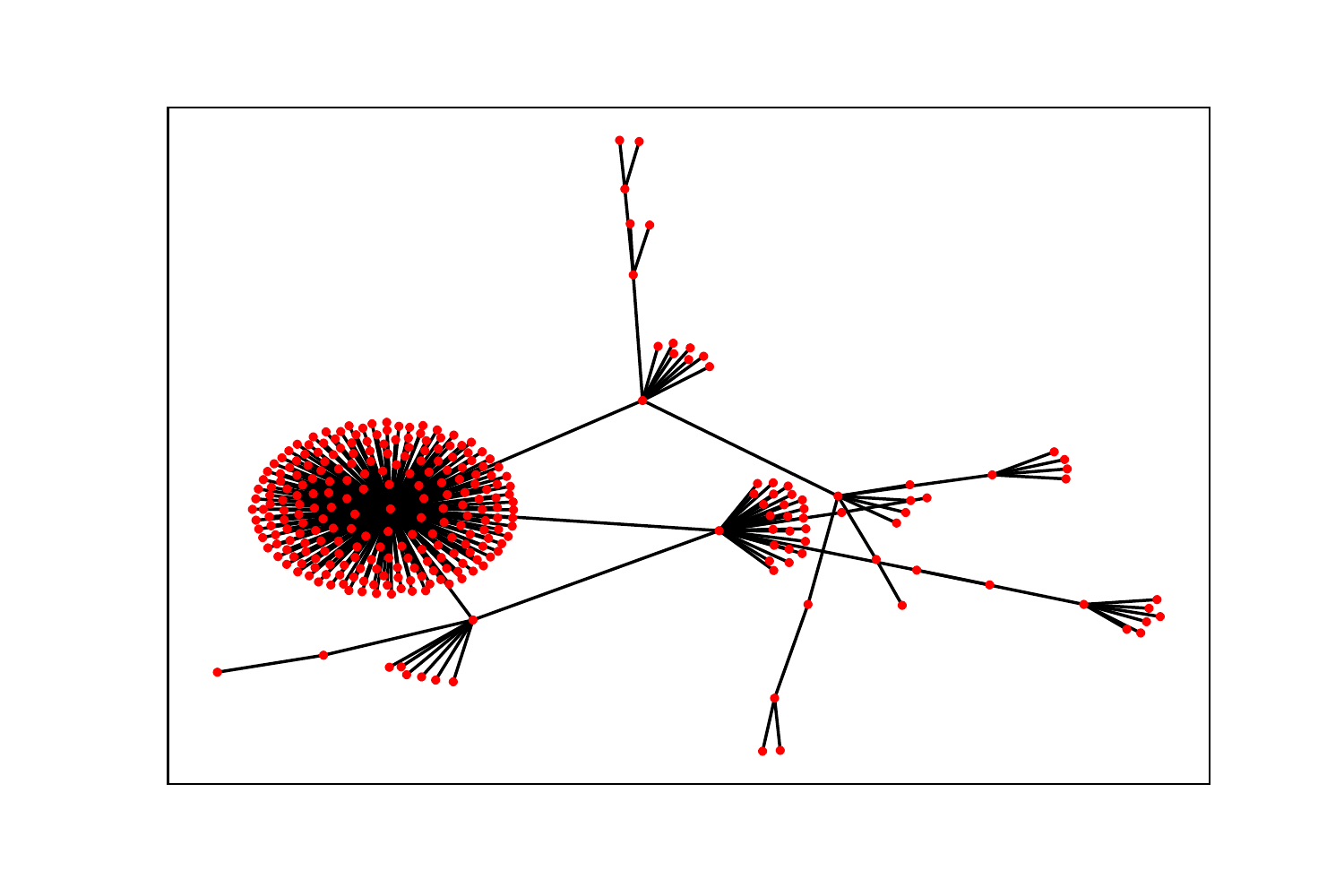}
			\label{fig:N5}
		\end{subfigure}
		\hfill
		\begin{subfigure}[b]{0.49\textwidth}
			\centering
			\subcaption[textfont=large]{}
			\includegraphics[width=\textwidth]{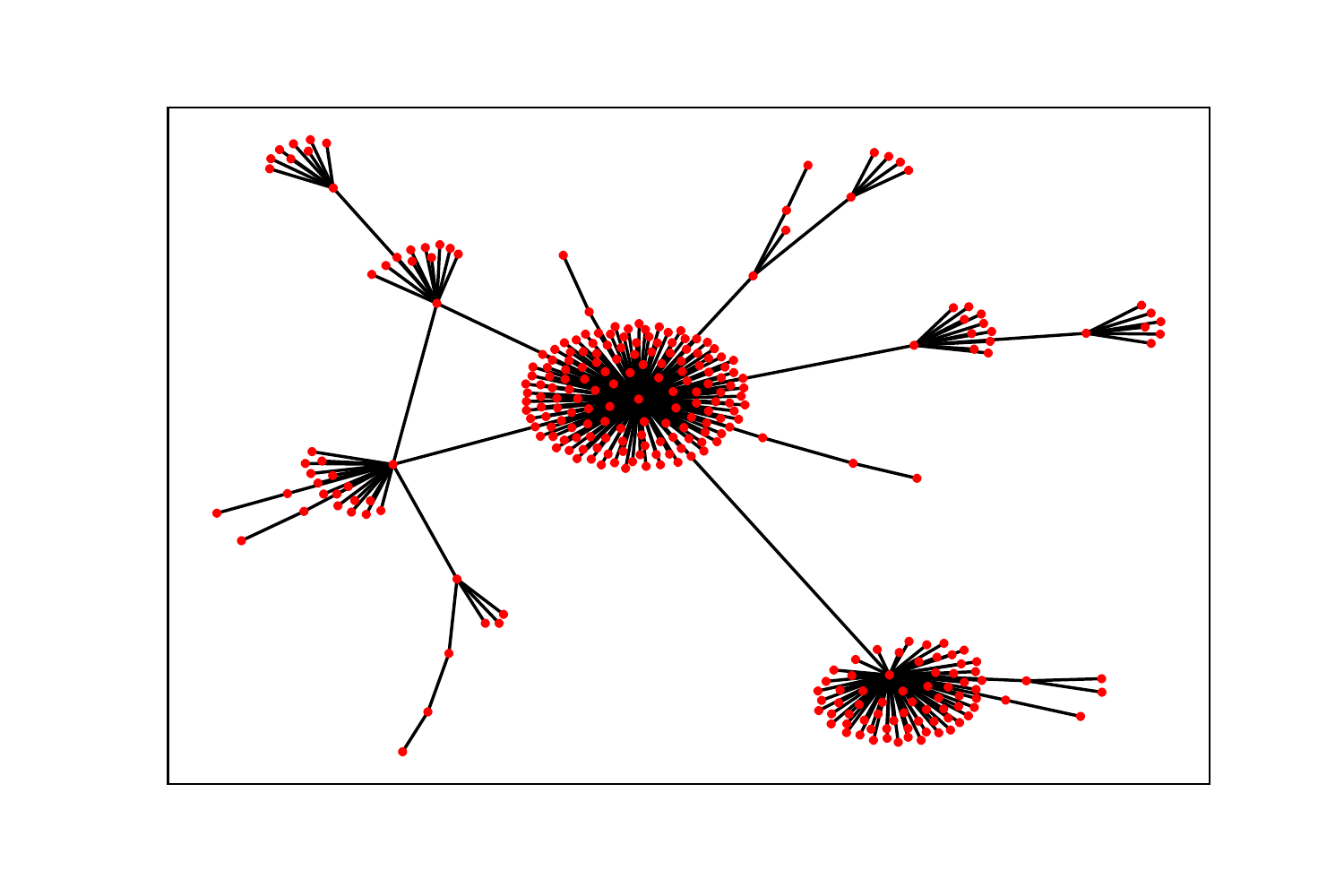}
			\label{fig:N6}
		\end{subfigure}
	\vspace*{-18mm} 
		\caption{ \textbf{Visualization of networks produced with $p= 0$ and $q =1$}. Each network diagram is generated with a different run but the same size $t = 300$}
		\label{Network Visualization}
	\end{figure}

The way to answer this question is to look at whether or not the degree distribution is stationary (see 'related models' for background on non-stationary distributions \cite {Gabel2013, Gabel2014, Lambiotte, Bhat, Krapivsky2017}). To this end, we started by visualizing six network  topologies, each generated with only $300$ nodes, with $p=0$ and $q=1$. Fig. \ref{Network Visualization} shows that diverse network structures arise, from tree-like (Figures \ref{Network Visualization}A, \ref{Network Visualization}B, \ref{Network Visualization}E, and \ref{Network Visualization}F) to star-like networks (Figures \ \ref{Network Visualization}B and \ref{Network Visualization}D). They are, in general, characterised by one or more central hubs to which the majority of nodes are connected. The emergence of these structures can be explained heuristically, as follows. At the start of the simulation, we have three connected nodes. Initially, and quite quickly, one of these (or another of the nodes added near the start) becomes a hub: attracting more connections than the other. From then on, the added nodes will only have one attachment, i.e.\ to the hub, while the hubs are attached to a large number of nodes. For example, imagine the situation where we have 20 nodes, 2 of which are hubs and the others, non-hubs, have only one connection, directly to a hub. In this case, the probability that when we make the first random choice (i.e.\ the random target node) in the model that we choose a non-hub is 9 in 10 (since only 1 in 10 of the nodes are hubs). Then, since all non-hubs are attached to hubs, our new node will also attach to a hub (i.e.\ the non-hub has only one neighbour, so the random neighbour must be a hub). As a result, hubs increase their degree even further and are thus even more likely to be attached to. Far from being standard preferential attachment, where attachment is proportional to the number of edges, this attachment disproportionally prefers hubs over non-hubs. 

Returning to our original derivation of the master equation (in particular, Eq.\ (\ref{BayesBC})), with an assumption that $B$ (i.e. the random target node) was equally likely to be friends with any person in the network, leading us, in Eq.\ (\ref{BayesBC}) to assume that the probability $B$ is friends with $C$ is proportional to $k$. While this assumption appears to hold when $p$ is close to one, because there are lots of random attachments in the network, the simulations where $p=0$ and $q=1$ show that this assumption is wrong. Indeed, the probability that $B$ is friends with a non-hub is very small. 

\begin{figure}
	\centering
	\hspace*{-8mm} 
	\includegraphics[width=19cm]{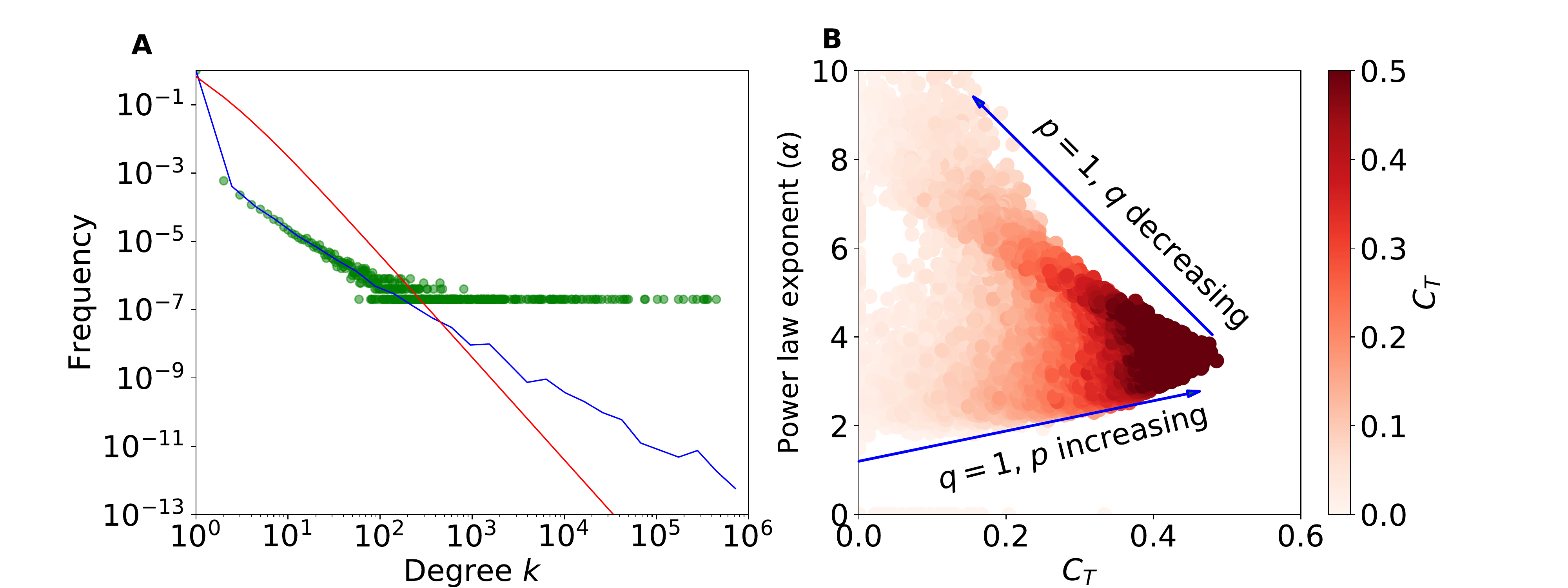}
	\vspace*{-20mm} 
	\caption{Panel A shows numerical and theoretical degree distributions for a network with $t=10^{6}$. The green scattered points and blue line represent the numerical simulations and the red line represent incorrect predicted values. The numerical simulations are averaged over 5 runs. The simulation is parameterized with $p = 0$ and $q =1 $. Panel B shows power-law exponents versus clustering coefficients for a total of 101 networks each with 100,000 nodes parameterized with different values of $p$ and $q$}
	\label{Expo_Vs_Clu_combined_LoglogplotPDF_p_zero}
\end{figure}

To investigate what type of distributions arise when $p$ is close to zero and $q$ is close to one, we simulated the model and measured the properties of the resulting distribution. First we simulated the $p =0$ and $q = 1$ model with $t=1000,000$ nodes, 5 times. The average frequency of degree over all 5 simulations is shown as the blue line   in Fig.\ \ref{Expo_Vs_Clu_combined_LoglogplotPDF_p_zero}A, while the degree distribution for each simulation that appeared is shown by the scattered green points. Computing the power-law exponent by the maximum likelihood method proposed by Newman \cite{Newman}, with a lower cut-off value $k=10$, gave $\alpha=1.61$. This is very different from the value of $\alpha = 3$  obtained from theoretical predictions (shown as the red line in Fig. \ref{Expo_Vs_Clu_combined_LoglogplotPDF_p_zero}A).  They are however, relatively consistent with those of Krapivsky and Redner \cite{Krapivsky2017} who, as discussed in section 2, found that the $p=0$ and $q =1$ generates non-stationary networks with super-hub nodes. They estimated an exponent $\alpha \approx 1.566$. 

Some care is required in how we interpret the power-law fitting of $\alpha$. Due to the effects of the condensation phenomenon \cite{Bianconi2001, Dorogo2001, Dorogo2002} and as we already saw in Fig.\ \ref{Network Visualization}, the hub-like networks resulting from this model do not resemble the types of the network we would expect from a power-law distribution. Moreover, when we look at the scatter of green points in the distribution of degrees for each of the individual simulations, we find a very large variance in the number of connections going into the most connected hub. This suggests that, despite the power-law fit to the average of a simulation where the number of nodes is finite, that in the limit of a large network the distribution is not a power-law. 

To demonstrate this last statement, we derived a new master equation, just for the model where $p=0$ and $q=1$, and concentrated on looking at how the proportion of nodes with degree $k=1$ changed over time. Firstly, note that the new nodes added to the graph are always degree 1 initially. Secondly, the only way in which the number of degree $k=1$ nodes can decrease is if the first node (what we have called $B$ in Fig.\ \ref{Network formation}) does not have degree 1, and then the second node we choose (what we have called $C$ in Fig. \ref{Network formation}) does have degree 1. This leads to the following master equation
\begin{equation}  \label{EquationNewMaster}
 	(t +1)P_{1}(t + 1) =  tP_1(t) + 1 - \left(1-P_1(t) \right) P_C(t)
\end{equation}
where $P_1(t)$ is the proportion of nodes with degree $k=1$. The undefined quantity in this equation is $P_C(t)$, which is the probability that the second node ($C$) chosen has degree $k=1$, given that the first node ($B$) chosen does not, i.e.
$P_C(t)=P(K_C=1 | K_B >1)$.

In Appendix B we prove that $P_C(t) \leq P_1(t)$. The key to this proof is to show, using Lagrange multipliers, that the case where the hubs (in this case defined as all nodes which have $k>1$) have an equal number of connecting non-hubs (i.e. nodes with $k=1$). In this case, if $m$ is the number of hubs and $t-m$ is the number of non-hubs, then 
\[
P_C(t) \leq \frac{\frac{t-m}{m}}{1+\frac{t-m}{m}} =  \frac{t-m}{t} = P_1(t) 
\]
Given that $P_C(t) \leq P_1(t)$, we can define a new master equation
 \begin{equation}  \label{EquationNewMasterBound}
 	(t +1)P^*_{1}(t + 1) =  tP^*_1(t) + 1 - \left(1-P^*_1(t) \right) P^*_1(t) 
\end{equation}
that has the property that $P^*_{1}(t) \leq P_{1}(t)$ for all $t$. Now, as $t \rightarrow \infty$, 
 \begin{equation}  \label{EquationNewMasterBound1}
 	P^*_{1} =  1 - \left(1-P^*_1 \right) P^*_{1}
\end{equation}
which has solution $P^*_{1} =1$. Thus the number of non-hubs goes to $1$ in the limit of large $t$. Since $P^*_{1}(t) \leq P_{1}(t)$ it thus follows that Eq. (\ref{EquationNewMaster}) also goes to $1$. In an infinitely large graph, all but a finite number of nodes, become non-hubs.   

Real-world graphs are, of course, finite. So it is still relevant to look at how clustering and the empirical power-law exponent depend on the parameters $p$ and $q$. To do this, we simulated the model for $t=100,000$ nodes for different values of $p$ and $q$. Fig.\ \ref{Expo_Vs_Clu_combined_LoglogplotPDF_p_zero}B shows the values of $C_T$ and $\alpha$ obtained. The red color intensity in Fig.\ \ref{Expo_Vs_Clu_combined_LoglogplotPDF_p_zero}B is proportional to  the product of $p$ and $q$, highlighting that a high clustering coefficient and $\alpha$ values close to around $3.47$ occur when both $p$ and $q$ are close to $1$. When $q$ approaches 0 then $\alpha$ increases, while when $p$ approaches 0 then $\alpha$ decreases. This gives a feasible region for the clustering $C_T$ (of between $0$ and $0.47$) and an empirically measured power-law exponent $\alpha$ (of between about $1.5$ and $\infty$). 

	\section{Discussion}

We have investigated the friend of a friend model for network formation and found a surprisingly rich diversity of resulting networks results. The mechanism behind our model is both local and simple. We would claim that it is difficult to find a simpler mechanism: connections are made by first picking one person at random and then picking one of their friends at random. Yet, at a global network level, this model produces many properties which are considered very important for complex networks. By just varying $p$ and $q$, we can simulate networks spanning from random networks through scale-free networks to super-hub networks.  These networks can have average clustering coefficients ranging from $0$ to $0.47$ and exhibit small-world characteristics.
	
	It is impossible to reconcile our results with some of the previous claims in the literature about complex networks. It appears that in the first articles investigating the model as we study here, the authors did not concern themselves closely enough looking into the discrepancy between numerical models and the mean-field approximation they made \cite{ Jackson}. Instead, they claimed that the $p = 0$ and $q = 1$ model had the same mean-field model, and thus a power-law exponent of $3$, as the Barab\'asi and Albert's preferential attachment model \cite {BA}.  Similar claims are repeated in Barab\'asi's {\it Network Science} book \cite {BAbook}, in a statement regarding the copying models. Both our numerical simulations and the proof that the proportion of nodes with degree $k=1$ tends to $1$ contradict these claims.
	
We believe that our model should prove useful help in navigating data collected on complex networks. There are still ongoing discussions on how scale-free networks must be characterized \cite{Broido, HolmeP,Liu}. In the context of our model, we would simply encourage researchers to measure the slope of the empirical power-law $\alpha$ in their data together with the average clustering co-efficient $C_T$ and see if it lies within the red shaded area in Fig.\ 	\ref{Expo_Vs_Clu_combined_LoglogplotPDF_p_zero}B. Power laws with empirical exponents as large as of about $1.5$ are possible in our model, as are average clustering coefficients up to around $0.5$. It is thus possible to identify (reverse engineer) the values of $p$ and $q$ consistent with empirical measurements of clustering and power law exponent. The friend of a friend model with those values of $p$ and $q$ thus becomes a candidate mechanism for generating the network.  

It is also worth pausing to think about our model in the context of the idea that the rich get richer. Often preferential attachment is associated with this concept, under which individuals who are seen by others to have social resources (i.e. friends) are more likely to  attract more attachments. Power-laws, or long-tail distributions, are associated with the rich getting richer. The friend of a friend model generates both small-world clustering and power-laws without individuals explicitly preferring to attach to 'richer' individuals. It is the recommendation of others that leads to the power-law. In the most extreme variation of the model ($p=0$ and $q=1$), the distribution is even more skewed towards individuals who happened to attract the most attachments initially.

	\section*{Appendix A}
\section*{Mean-field derivation of the degree distribution $P_k$}

Here, we make a theoretical derivation of the degree distribution $P_k$ using the mean-field analysis. 

 Initially at time step $t = t_i$, a new node with  an average of $k(t_i) = E[K_C]$ new degrees is added. 
 In one time step, the total rate of increase of degree $k$ is given by 
 \begin{equation} \label{rateequation}
 	\frac{dk}{dt}=  \frac{p}{t} + q \frac{k}{2(p+q)t}. 
 \end{equation} 
 
The solution of Eq. (\ref{rateequation}), gives the prediction of the degree distribution of the network, 
\begin{equation} \label{solution1}
	k(t) + \frac{2p(p+q)}{q} = \left(2(p+q) +\frac {2p(p+q)}{q}\right)\left(\frac{t}{t_i}\right)^\frac{q}{2(p+q)}.
\end{equation} 

Eq. (\ref{solution1}) can also be written as, 
\begin{equation} \label{solution2}
	t_i =\frac{\left(2(p+q) +\frac {2p(p+q)}{q}\right)^\frac{2(p+q)}{q}} {\left(k +\frac {2p(p+q)}{q}\right)^\frac{2(p+q)}{q}}t.
\end{equation}
Suppose that new nodes are added to the network at equal time interval. The probability density function for any time $t_i$ falling within this particular time interval [$t_i, t$] may be assumed to be,

\begin{equation} \label{pdf}
	P_{k}(t)=\frac{1}{N}=\frac{1}{2(p+q) + t}.
\end{equation}
Therefore, the probability that a node has degree  $k(t_i)$ smaller than $k(t)$ (cummulative degree distribution) can be written as 

\begin{equation} \label{pdf1}
	P \left(k(t_i)<k\right) = \frac{(2(p+q) +t)-t_i} {2(p+q)+t} = 1-\frac{t_i} {2(p+q)+t}.
\end{equation}

Substituting Eq. (\ref{solution2}) into Eq. (\ref{pdf1}) we have
\begin{equation} \label{pdf2}
	P \left(k(t_i)<k\right) = 1-\frac{\left(2(p+q) +\frac {2p(p+q)}{q}\right)^\frac{2(p+q)}{q}t} {(2(p+q)+t){\left(k +\frac {2p(p+q)}{q}\right)^\frac{2(p+q)}{q}}}.
\end{equation}
The probability density function (degree distribution) $P\left( k(t)\right) $ is given by the partial derivative of Eq. (\ref{pdf2}) with respect to $k$. Thus,

\begin{equation} \label{pdf3}
	P\left( k(t)\right) =\frac{\partial P \left(k(t_i)<k\right)}{\partial k}= \frac{2(p+q)}{q} \frac{\left(2(p+q) +\frac {2p(p+q)}{q}\right)^\frac{2(p+q)}{q}t} {(2(p+q)+t){\left(k +\frac {2p(p+q)}{q}\right)^{1+\frac{2(p+q)}{q}}}}.
\end{equation}
As ${t \to \infty}$, $P\left( k(t)\right)$  gets closer and closer to 
$P\left( k\right)$. Therefore,
\begin{equation}\label{pdf4}
	P\left( k\right) = \frac{2(p+q)}{q} \left(2(p+q) +\frac {2p(p+q)}{q}\right)^\frac{2(p+q)}{q} \left(k +\frac {2p(p+q)}{q}\right)^{-(1+\frac{2(p+q)}{q})}.
\end{equation} 

In the limits of very large $k$, the degree distribution exhibits the extended power-law given by 
\begin{equation}\label{pdf6}
	P\left( k\right) \sim\left(k +\frac {2p(p+q)}{q}\right)^{-(1+\frac{2(p+q)}{q})}  \sim k ^{-(1+\frac{2(p+q)}{q})} , \quad \text{for all} \quad k.
\end{equation} 
From Eq. (\ref{pdf6}), the scaling parameter $\alpha$ can be computed using the exponent of $k$, 

\begin{equation}\label{alpha}
	\alpha = 1+\frac{2(p+q)}{q}=3 +\frac {2p}{q}.
\end{equation} 
The value of $\alpha$ must always fall within the interval $[3, +\infty)$.
When $p=0$ and $q = 1$, the model predicts $\alpha =3$.\\

The master equation for directed networks is similarly derived, but
with $E[K_C] = p+q$. This gives, $\alpha = 2 +\frac {p}{q}$.

	\section*{Appendix B}
\section*{Proportional of nodes with degree $k=1$ when $t\to \infty$}

In this section we prove that $P_C(t) \leq P_1(t)$. \\

Consider a graph that has $m$ hubs (nodes with degree greater than one) on time step $t$. The number of non-hubs (degree one) is thus $t-m$. We denote the number of non-hubs connected to a hub as $n_i$ and note that
\begin{equation}
n_1 + n_2 + ... + n_m = t-m \label{PNcon}.
\end{equation}
If the first node chosen (i.e. B) is a non-hub, then the probability ($P_C(t)$) that we choose a hub is given by
\[
1 - P_C(t) = \frac{1}{m} \left( \frac{h_1}{n_1 +h_1} + \frac{h_2}{n_1 +h_2}  + ... + \frac{h_m}{n_m +h_m} \right),
\]
where $h_i \geq 1$ is the number of hubs connected to hub $i$. We then note that 
\begin{equation}
1 - P_C(t) \geq \frac{1}{m} \left( \frac{1}{n_1 +1} + \frac{1}{n_2 +1}   + ... + \frac{1}{n_m + 1} \right). \label{PNmin}
\end{equation}
We then define a Lagrange multiplier, such that
\[
L = \frac{1}{m} \left( \frac{1}{n_1 + 1} + \frac{1}{n_2 +1}   + ... + \frac{1}{n_m + 1} \right) + \lambda \left( n_1 + n_2 + ... + n_m - t + m \right),
\]
which allows us to minimise the expression in Eq. (\ref{PNmin}) subject to constraint Eq. (\ref{PNcon}) by solving
\[
\frac{\partial L}{\partial n_i} = - \frac{1}{(n_i + 1)^2}  + \lambda= 0
\]
to give $\lambda= \left( m/(t-m) + 1 \right)^2$ and
\[
\frac{\partial L}{\partial \lambda} = n_1 + n_2 + ... + n_m - t + m = 0,
\]
which, after substituting in $\lambda$ gives
\[
n_i = \frac{t-m}{m}.
\]
This means that the expression in Eq. (\ref{PNmin}) is minimised when the non-hubs are evenly distributed across the hubs. We thus see that,
\begin{equation}
1 - P_C(t) \geq \frac{1}{m} \left( \frac{m}{ \frac{t-m}{m}+1} \right) = \frac{m}{t} = 1 - P_1(t)
\end{equation}
Therefore, $P_C(t) \leq P_1(t)$, as required.

\section*{Funding}	
This research work received no specific funding.
	
\section*{Acknowledgements}

We thank Petter Holme for inputs and comments and Anton V. Proskurnikov for his time and fruitful discussion.

\newpage
	
	
\end{document}